\documentstyle[epsf,times]{mn2e}

\newcommand{\simgt}{\lower.5ex\hbox{$\; \buildrel > \over \sim \;$}}
\newcommand{\simlt}{\lower.5ex\hbox{$\; \buildrel < \over \sim \;$}}
\newcommand{\bm}[1]{\mbox{{\it \boldmath$#1$}}}
\newcommand{\kaco}[1]{\left\langle{#1}\right\rangle}
\newcommand{\skaco}[1]{\langle{#1}\rangle}

\newcommand{\apj}{ApJ}

\newcommand{\mnras}{MNRAS}
\newcommand{\LCDM}{$\Lambda$CDM~}

\newcommand{\rvir}{r_{\rm vir}}

\newcommand{\baredth}{\;\overline{\raise1.0pt\hbox{$'$}\hskip-6pt
\partial}\;}
\newcommand{\edth}{\;\raise1.0pt\hbox{$'$}\hskip-6pt\partial\;}

\begin{document}
\title[Magnification statistics]{Halo model predictions of the cosmic
magnification statistics: the full non-linear contribution}

\author[M. Takada \& T. Hamana]
{
Masahiro Takada$^1$\thanks{E-mail: mtakada@hep.upenn.edu}
and Takashi Hamana$^{2,3}$\thanks{E-mail: hamana@iap.fr} \\
$^1$ Department of Physics
and Astronomy, University of Pennsylvania, 209 S. 33rd Street,
Philadelphia, PA 19104, USA\\
$^2$ Institut d'Astrophysique de Paris, 98 bis Bld Arago, F-75014,
Paris, France\\
$^3$ National Astronomical Observatory of Japan, Mitaka, Tokyo
181-8588, Japan
}

\pagerange{\pageref{firstpage}--\pageref{lastpage}}

\maketitle

\label{firstpage}
\begin{abstract}
The lensing magnification effect due to large-scale structure
is statistically
measurable by correlation of size fluctuations in distant galaxy images
as well as by cross-correlation between foreground galaxies and background
sources such as the QSO-galaxy cross-correlation.
We use the halo model formulation of Takada \& Jain (2003) to compute
these magnification-induced correlations without employing
 the weak lensing approximation, $\mu\approx 1+2\kappa$.
Our predictions thus include the full contribution from non-linear
magnification, $\delta \mu\simgt 1$, that is due to lensing
halos.  We compare the model prediction with ray-tracing simulations and
find excellent agreement over a range of angular scales we consider
($0.\!\!'5\la \theta \la 30'$). In addition,  we derive
the dependence of the correlation amplitude on 
the maximum magnification cutoff $\mu_{\rm max}$, 
which is necessary to introduce in
 order to avoid the contributions from strong lensing events. 
For a general correlation function parameterized as
$\skaco{\mu^p f}$ ($f$ is any cosmic field correlated with the
magnification field), the amplitude remains finite for $p<1$ and diverges for
$p\ge 1$ as $\mu_{\rm max}\rightarrow \infty$, independent of details of the
lensing mass distribution and of the separation angle.  This consequence is
verified by the halo model as well as by the simulations.
Thus the magnification correlation with $p\le 1$ has a practical advantage
in that it is insensitive to a selection effect of how strong lensing
events with $\mu\gg 1$ are observationally excluded from the sample.

The non-linear magnification contribution
enhances the amplitude of the magnification correlation relative to the
 weak lensing approximation, and 
the non-linear correction
is more significant on smaller angular scales and for sources at higher
redshifts.  
The
enhancement amounts to $10-25\%$ on arcminite scales for the QSO-galaxy
cross-correlation, even after inclusion of a realistic model of galaxy
clustering within the host  halo.
Therefore, it is necessary  to account for the
non-linear contribution in theoretical models in order to make an
unbiased, cosmological 
interpretation of the precise measurements expected from
forthcoming massive surveys. 
\end{abstract}

\section{Introduction}

Gravitational lensing caused by the large-scale structure is now
recognized as a powerful cosmological tool (Mellier 1999; Bartelmann \&
Schneider 2001 for thorough reviews).  The gravitational deflection of
light causes an increase or decrease of the area of a given patch on the
sky depending on which the light ray passes preferentially through the
overdense or underdense region. Accordingly, this causes an observed
image of source to be magnified or de-magnified relative to the unlensed
image, since lensing conserves the surface brightness and the received
luminosity is proportional to the solid angle of the image. Large
magnifications are observed in a strong lensing system 
that accompanies multiple images or largely deformed images.  It has been
also proposed that mild or weak magnifications is measurable in a statistical
sense. The magnification leads to an enhancement in the flux-limited
number counts of background sources around foreground sample that traces
the lensing mass distribution.  Based on this idea, numerous works have
investigated the QSO-galaxy cross correlation theoretically (e.g.,
Broadhurst, Taylor \& Peacock 1995; Bartelmann 1995) as well as
observationally (e.g., Benitez \& Mart\'inez-Gonz\'alez 1997; Benitez et
al. 2001; Gazta\~naga 2003; also see Bartelmann \& Schneider 2001 for a
thorough review). Further, 
Jain (2002) recently proposed that the magnification
effect can be extracted by statistically dealing with size fluctuations
of distant galaxy images, though it is a great challenging for existing
data yet.  Forthcoming massive surveys such as SDSS\footnote{\tt
http://www.sdss.org/}, DLS\footnote{\tt dls.bell-labs.com/}, the CFHT
Legacy survey\footnote{\tt www.cfht.hawaii.edu/Science/CFHLS/} as well
as future precise imaging surveys such as SNAP\footnote{\tt
snap.lbl.gov}, Pan-STARRS\footnote{\tt www.ifa.hawaii.edu/pan-starrs/}
and LSST\footnote{\tt www.dmtelescope.org/dark\_home.html} allow
 measurements of these magnification effects at high significance.
 Therefore, it is of
great importance to explore how this type of method can be a useful
cosmological tool and complementary to the established cosmic shear
method, which measures correlation between lensing induced ellipticities
of distant galaxy images (e.g., Hamana et al. 2002 and Jarvis et
al. 2003 and references therein).

The magnification field $\mu(\bm{\theta})$ in a given direction
$\bm{\theta}$ on the sky is expressed (e.g, Schneider, Falco \& Ehlers
1992) as
\begin{equation}
\mu(\bm{\theta})=\left|
(1-\kappa(\bm{\theta}))^2-\gamma^2(\bm{\theta})\right|^{-1}.
\label{eqn:mag}
\end{equation}
Here $\kappa$ and $\gamma$ are the convergence and shear fields, which
are fully determined by the mass distribution along the line of sight.
This equation shows the nonlinear relation between the magnification and
the convergence and shear, and indicates that the magnification
increases with $\kappa$ and $\gamma$ very rapidly and becomes even
formally infinite when the lensing fields $\kappa,\gamma\sim O(1)$.
However, as
long as we are concerned with the magnification related statistics due
to the
large-scale structure, strong lensing events ($\mu\gg 1$) should be
removed from the sample to prevent the large statistical scatters. 
This will be straightforward to implement, if
the strong lensing accompanies multiple images or largely deformed
images. On the other hand, 
modest magnification events ($\delta\mu\simgt 1$) make it relatively
difficult to identify and are likely included in the sample for the
blind analysis, 
since the magnification is not a direct observable. Therefore, the
magnification statistics rather requires a more careful study of the selection
effect than the cosmic shear (e.g., Barber \& Taylor 2003), which we
will carefully address. 

The simplest statistical quantity most widely used in cosmology is the
two-point correlation function (2PCF).  For our purpose, the
magnification field is taken as either or both of the two fields
entering into the correlation function.  However, it is not
straightforward to analytically compute the magnification 2PCF because
of the non-linear relation between $\mu$ and $\kappa,\gamma$, where the
latter fields are easier to compute in a statistical sense based on a
model of the mass power spectrum.  For this reason, the conventional
method of the magnification statistics employs the weak lensing
approximation $\mu\approx 1+2\kappa$ (e.g., Bartelmann 1995; Dolag \&
Bartelmann 1997; Sanz et al. 1997; Benitez \& Mart\'inez-Gonz\'alez
1997; Moessner \& Jain 1998; Benitez et al. 2001; M\'enard \& Bartelmann
2002; Gazta\~naga 2003; Jain et al. 2003).  However, it is
obvious that this type of method is valid only in the limit
$\kappa,\gamma\ll1$ and likely degrades the model accuracy on non-linear
small scales.  
In fact, using the ray-tracing simulations M\'enard et
al. (2003) clarified the importance of the non-linear
magnification contribution to the magnification statistics (see also
Barber \& Taylor 2003).  It was shown that the perturbative treatment
breaks down over a range of angular scales of our interest.  Although a
promising method to resolve this issue 
is to employ ray-tracing simulations, to
perform multiple evaluations in model parameter space requires
sufficient number of simulation runs, which is relatively prohibitive.

Therefore, the main purpose of this paper is to develop an analytic method to
compute the magnification induced correlation function without employing
the weak lensing approximation.
To do this, we use the halo model to describe gravitational clustering
in the large-scale structure, following the method developed in Takada
\& Jain (2003a,b,c, hereafter TJ03a,b,c). 
The model prediction of the QSO-galaxy
cross-correlation is also developed 
by incorporating a realistic model of galaxy
clustering within the host halo into the halo model.  
Although the halo model rather
relies on the simplified assumptions, 
the encouraging results revealed so far are that
it has led to consistent
predictions to interpret observational results of galaxy clustering as
well as to reproduce simulation results (e.g., see Seljak 2000; Zehavi et
al. 2003; Takada \& Jain 2002, hereafter TJ02; TJ03a,b,c; also see
Cooray \& Sheth 2002 for a review).

Another purpose of this paper is to explore how the magnification related
statistics can probe the halo structure.  The non-linear
magnifications arise when the light ray
emitted from a source encounters an intervening mass concentration,
i.e., dark matter halo such as galaxy or cluster of galaxies.  
It is known that strong lensing of $\mu\gg 1$ can be used to probe
detailed mass distribution within a halo (e.g., Hattori et al. 1999).  
Similarly, modest non-linear magnifications of $\delta\mu\simgt 1$
could lead to a sensitivity of the magnification statistics to the halo
structure in a statistical sense, as investigated in this paper.  A
fundamental result of cold dark matter (CDM) model simulations is that
the density profiles of halos are universal across a wide range of mass
scales (e.g., Navarro, Frenk \& White 1997, hereafter NFW). On the other
hand, some alternative models such as self-interacting dark matter
scenario (Spergel \& Steinhardt 2000) have been proposed in order to
reconcile the possible conflicts between the simulation prediction and
the observation.  
If dark matter particle has a
non-negligible self-interaction between themselves, the effect is likely
to yield a drastic change on the halo profile compared to the CDM model
prediction (Yoshida et al. 2000). The halo structure thus reflects the
dark matter nature as well as detailed history of non-linear
gravitational clustering.  Hence, exploring the halo profile properties
with gravitational lensing can be a direct test of the CDM paradigm on
scales $\simlt $Mpc, which is not attainable with the cosmic microwave
background measurement.

The plan of this paper is as follows. \S \ref{prel} presents the basic
equations relevant for cosmological gravitational lensing and then
briefly summarize two promising methods to statistically measure the
magnification effect.  In \S \ref{model} we develop an analytic method
to compute the magnification correlation functions based on the halo
model. In \S \ref{asymp}, we derive an asymptotic behavior of the
correlation amplitude for large magnifications. In \S \ref{results} we
qualitatively test the halo model prediction and the asymptotic
behavior using ray-tracing simulations. The realistic model of the
QSO-galaxy cross-correlations is also derived. 
 \S \ref{disc} is devoted to a
summary and discussion.  Throughout this paper, without being explicitly
stated, we consider the $\Lambda$CDM model with $\Omega_{m0}=0.3$,
$\Omega_{\lambda0}=0.7$, $\Omega_{\rm b0}$, $h=0.7$ and $\sigma_8=0.9$.
Here $\Omega_{\rm m0}$, $\Omega_{b0}$ and $\Omega_{\lambda0}$ are the
present-day density parameters of matter, baryons and the cosmological
constant, $h$ is the Hubble parameter, and $\sigma_8$ is the rms mass
fluctuation in a sphere of radius $8h^{-1}$Mpc.

\section{Preliminaries}
\label{prel}

\subsection{Magnification of gravitational lens}

The gravitational deflection of light ray induces a mapping between
the two-dimensional source plane (S) and the image plane (I) (e.g., 
Schneider, Ehlers \& Falco 1992):
\begin{equation}
\delta\!\theta_i^{\rm S}={\cal A}_{ij}\delta\!\theta_J^{\rm I},
\label{eqn:map}
\end{equation}
where $\delta\theta_i$ is the separation vector between points on the
respective planes.  The lensing distortion of an image is described by
the Jacobian matrix ${\cal A}_{ij}$ defined as
\begin{equation}
{\cal A}=
\left(
\begin{array}{cc}
1-\kappa-\gamma_1 & -\gamma_2\\
-\gamma_2 & 1-\kappa+\gamma_1
\end{array}
\right),
\end{equation}
where $\kappa$ is the lensing convergence and $\gamma_1$ and $\gamma_2$
denote the tidal shear fields, which correspond to elongation or
compression along or at $45^\circ$ to $x$-axis, respectively, in the
given Cartesian coordinate on the sky.  The $\kappa$ and $\gamma_i$
depend on angular position, although we have omitted showing it in the
argument for simplicity.  Since the gravitational lensing conserves
surface brightness from a source, the lensing magnification, the ratio
of the flux observed from the image to that from the unlensed source,
is given by determination of the deformation matrix, yielding equation
(\ref{eqn:mag}).
In the weak lensing limit $\kappa,\gamma\ll1$, one can Taylor expand
the magnification field as $\mu\approx 1+2\kappa$, as conventionally
employed in the literature to compute the magnification statistics.  The
weak lensing approximation ceases to be accurate as
$\kappa,\gamma\rightarrow O(1)$. 
  For example, $\kappa=0.5$ (and
simply $\gamma=0$) leads to factor 2 difference as $\mu=4$ and
$1+2\kappa=2$.

In the context of cosmological gravitational lensing, the convergence
field is expressed as a weighted projection of the three-dimensional
density fluctuation field between source and observer:
\begin{equation}
\kappa(\bm{\theta})=\int_0^{\chi_H}\!\!d\chi W(\chi) 
\delta[\chi, d_A(\chi)\bm{\theta}],
\label{eqn:kappa}
\end{equation}
where $\chi$ is the comoving distance, $d_A(\chi)$ is the comoving
angular diameter distance to $\chi$, and $\chi_H$ is the distance to the
Hubble horizon.  Note that $\chi$ is related to redshift $z$ via the
relation $d\chi=dz/H(z)$ ($H(z)$ is the Hubble parameter at epoch $z$).
Following the pioneering work done by Blandford et al. (1991),
Miralda-Escude (1991) and Kaiser (1992), we used the Born approximation,
where the convergence field is computed along the unperturbed path, 
neglecting higher order terms that arise from coupling between two or
more lenses at different redshifts.  Using ray-tracing simulations of
the lensing fields, Jain et al. (2000) proved that the Born
approximation is a good approximation for lensing statistics (see also
Van Waerbeke et al. 2001; Vale \& White 2003).  The lensing weight
function $W$ is given by
\begin{eqnarray}
W(\chi)&=&\frac{3}{2}\Omega_{m0}H_0^2 a^{-1}(\chi)
d_A(\chi)\nonumber\\
&&\times \int^{\chi_H}_\chi\!\!d\chi_s~ 
f_s(\chi_s) \frac{d_A(\chi_{\rm s}-\chi)}{d_A(\chi_s)},
\label{eqn:weightgl}
\end{eqnarray}
where $f_s(\chi_s)$ is the redshift selection function normalized as
$\int^{\chi_H}_0\!\! d\chi f_s(\chi)=1$. In this paper we assume all
sources are at a single redshift $z_s$ for simplicity;
$f_s(\chi)=\delta_D(\chi-\chi_s)$.  $H_0$ is the Hubble constant
($H_0=100h{~\rm km~s}^{-1}{\rm Mpc}^{-1}$). Similarly, the shear
fields are derivable from the density fluctuation fields, but the
relation is non-local due to the nature of the gravitational tidal
force. In Fourier space, we have the simple relation between 
$\kappa$ and $\gamma$: 
\begin{equation}
\tilde{\gamma}_1(\bm{l})=\cos2\varphi_{\bm{l}}\tilde{\kappa}(\bm{l}),
\hspace{1em}
\tilde{\gamma}_2(\bm{l})=\sin2\varphi_{\bm{l}}\tilde{\kappa}(\bm{l}),
\end{equation}
where the Fourier-mode vector is $\bm{l}=l(\cos\varphi_{\bm{l}},
\sin\varphi_{\bm{l}})$. 

\subsection{Methodology for measuring the magnification statistics}
\label{method}

There are two promising ways for measuring the lensing magnification
effect statistically, which are likely feasible for forthcoming and
future surveys. Here we briefly summarize the methodology.

Gravitational magnification has two effects. First, it causes the area
of a given patch on the sky to increase, thus tending to dilute the
number density observed. Second, sources fainter than the limiting magnitude are
brightened and may be included in the sample. The net effect, known as
the magnification bias, depends on how the loss of sources due to
dilution is balanced by the gain of sources due to flux magnification.
Therefore, it depends on the slope $s$ of the unlensed number counts of
sources $N_0(m)$ in a sample with limiting magnitude $m$, $s=d\log
N_0(m)/dm$.  Magnification by amount $\mu$ changes the number counts to
(e.g., Broadhurst, Taylor \& Peacock 1995; Bartelmann 1995);
\begin{equation}
N'(m)=N_0(m)\mu^{2.5s-1}.
\label{eqn:num}
\end{equation}
For the critical value $s=0.4$, magnification does not change the number
density; it leads to an excess for $s>0.4$, and a deficit for $s<0.4$.
Let $n_1(\bm{\theta})$ be the number density of foreground sample with
mean redshift $\skaco{z}_1$, observed in the direction $\bm{\theta}$ on
the sky, and $n_2(\bm{\theta})$ that of the source sample with a higher
mean redshift $\skaco{z}_2>\skaco{z}_1$.  Thus, even if there is no
intrinsic correlation between the two populations, magnification induces
the non-vanishing cross-correlation:
\begin{eqnarray}
\xi(\theta)&=&\skaco{\delta n_1(\bm{\theta}_1)\delta 
n_2(\bm{\theta}_2)}_{|\bm{\theta}_1-\bm{\theta}_2|=\theta}
\nonumber\\
&=&\skaco{\delta n_1(\bm{\theta}_1)
\left[\mu(\bm{\theta}_2)
\right]^{2.5s-1}}_{|\bm{\theta}_1-\bm{\theta}_2|=\theta},
\label{eqn:cross}
\end{eqnarray}
where $\delta n_i(\bm{\theta})=(n_i(\bm{\theta})-\bar{n}_i))/\bar{n}_i$,
with $\bar{n}_i$ the average number density of the $i$th sample. Here
$\skaco{\cdots}$ denotes the ensemble average and observationally means
the average over all pairs separated by $\theta$ on the sky.  Based on
this idea, numerous studies have confirmed the existence of the
enhancement of the QSO number counts around foreground galaxies, i.e.,
the QSO-galaxy cross correlation (e.g., Benitez et al. 2001, Gazta\~naga
2003 and references therein; also see Bartelmann \& Schneider 2001 for a
review).  Although these results are in qualitatively agreement with the
magnification bias, in most cases the amplitude of the correlation is
much higher than that expected from gravitational lensing models. The
excess might be due to the non-linear magnification contribution from
massive halos (e.g., see the discussion around Figure 28 in Bartelmann
\& Schneider 2001).  
However, to make such a statement with confidence, it is
necessary to further explore an obstacle in the theoretical model, the
bias relation between the galaxy and mass distributions.  This still
remains uncertain observationally and theoretically.  In particular, on
small scales ($\simlt $Mpc), it is crucial to model how to populate
halos of given mass with galaxies, known as the halo occupation number
(e.g., Seljak 2000; TJ03b). Recently, Jain, Scranton \& Sheth (2003)
carefully examined the effect of the halo occupation number on the
magnification bias and showed that the parameters used in it 
 yield a strong
sensitivity to the predicted correlation. For example,  possible
modification for types of galaxies leads to a change of a factor of 2-10
in the expected signal on arcminute scales. 
Further, we will show that the non-linear magnification correction is
also important to make the accurate prediction.

Second, Jain (2002) recently proposed a new method for measuring the
magnification effect of the large-scale structure, based on the work of
Bartelmann \& Narayan (1995). 
 The lensing effect is to increase or decrease an
observed image of galaxy relative to the unlensed image, depending on
which the light ray travels preferentially through overdense or
underdense region that corresponds to $\mu>1$ or $\mu<1$.  
That is, the area $S$ and characteristic radius
$R(\propto S^{1/2})$ are changed by the magnification field $\mu$ as
\begin{equation}
S\rightarrow S\mu, \hspace{2em} R\rightarrow R\mu^{1/2}.
\end{equation}
Although the unlensed image is not observable, this effect can be
statistically extracted as follows.  We can first estimate the mean size
of source galaxies from the average over the sample available from a given 
survey area, under the assumption $\skaco{\mu}=1$ or $\skaco{\mu^{1/2}}=1$.
Then, the two-point correlation function (2PCF)
of the size fluctuations can be computed from the average over all pairs
of galaxies separated by the angle considered, in analogy with the 2PCF
of the cosmic shear fields (the variance method was considered in Jain
2002).  The reason that this method is not yet feasible is that it
requires a well controlled estimate of the unlensed size distribution as
well as systematics (photometric calibration, resolution for sizes and
PSF anisotropy).  Hence, space based imaging surveys will make possible
the measurement of galaxy sizes with a sufficient accuracy hard to
achieve from the ground so far. This method can potentially be a 
precise cosmological probe as the cosmic shear measurement, because it
is free
from the galaxy bias uncertainty. Further, the non-linear nature of the
magnification could lead to complementarity to the cosmic shear
measurement, as we will discuss below.

\section{Halo Approach to magnification statistics}
\label{model}

\subsection{Halo profile and mass function}
\label{modeling} 

We use the the halo model of gravitational clustering to compute the
magnification statistics, following the method in TJ03a,b,c.  Key model
ingredients 
are the halo profile $\rho_h(r)$, the mass function $n(M)$, and the halo
bias $b(M)$, each of which is well investigated in the literature.

As for the halo profile, we employ an NFW profile truncated at radius
$r_{180}$, which is defined so that the mean density enclosed by sphere
with $r_{180}$ is $180$ times the background density.  Within a
framework of the halo model, we need to express the NFW profile in terms
of $M_{180}$ and redshift $z$.  To do this, we first express the two
parameters of the NFW profile, the central density parameter and the
scale radius, in terms of the virial mass and redshift, based on the
spherical top-hat collapse model and the halo concentration parameter of
Bullock et al. (2001) (see TJ03b,c for more details). Then, following Hu
\& Kravtsov (2002), we employ a conversion method between the virial
mass and $M_{180}$ in order to re-express the NFW profile in terms of
$M_{180}$ and $z$.  In what follows, we will often refer halo massed $M$
as $M_{180}$ for simplicity.

For the mass function, we employ the Sheth-Tormen mass function (Sheth
\& Tormen 1999):
\begin{eqnarray}
n(M)dM&=&\frac{\bar{\rho}_0}{M}f(\nu)d\nu\nonumber\\
&=&\frac{\bar{\rho}_0}{M}A[1+(a\nu)^{-p}]\sqrt{a\nu}\exp\left(-\frac{a\nu}{2}
\right)\frac{d\nu}{\nu},
\label{eqn:massfun}
\end{eqnarray}
where $\nu$ is the peak height defined by
\begin{equation}
\nu=\left[\frac{\delta_c(z)}{D(z)\sigma(M)}\right]^2.
\end{equation}
Here $\sigma(M)$ is the present-day rms fluctuation in the mass density,
smoothed with a top-hat filter of radius $R_M\equiv
(3M/4\pi\bar{\rho}_0)^{1/3}$, $\delta_c$ is the threshold overdensity
for the spherical collapse model and $D(z)$ is the linear growth factor
(e.g., Peebles 1980).  The numerical coefficients $a$ and $p$ are taken
from the results of Table 2 in White (2002) as $a=0.67$ and $p=0.3$,
which are different from the original values $a=0.707$ and $p=0.3$ in Sheth
\& Tormen (1999).  Note that the normalization coefficient $A=0.129$. 
The main reason we employ the halo boundary $r_{180}$
is that the mass function measured in $N$-body simulations can be better
fitted by the the universal form (\ref{eqn:massfun}) when one employs
the halo mass estimator of $M_{180}$, than using the virial mass
estimator, as carefully examined in White (2002).  To maintain the
consistency, we also employ the halo biasing $b(M)$ given in Sheth \&
Tormen (1999) with the same $a$ and $p$ parameters, which is needed for
the 2-halo term calculation. 
Note that $\int dM~b(M) n(M)M/\bar{\rho}_0\simeq 1$ for the halo bias
model and the mass function we employ (e.g., Seljak 2000).

\subsection{Lensing fields for an NFW profile}
\label{lensfield}

For an NFW profile, we can derive analytic expressions to give the
radial profiles of convergence, shear and magnification.

For a given source at redshift $z_s$, the convergence profile for a halo
of mass $M$ at $z$, denoted by $\kappa_M$, can be given as a function of
the projected radius from the halo center:
\begin{eqnarray}
\kappa_M(\theta)=4\pi Ga^{-1}\frac{d_A(\chi)d_A(\chi_s-\chi)}{d_A(\chi_s)}
\frac{M f}{2\pi r_s^2} F(\theta/\theta_{s}),
\label{eqn:conv}
\end{eqnarray}
where $r_{s}$ is the scale radius, $\theta_{s}$ its projected angular
scale and $f = 1/[\ln(1+c_{180})-c_{180}/(1+c_{180})]$ (see below for
the definition of $c_{180}$).  The explicit form of $F(x)$ is given by
equation (27) in TJ03b.  For an axially symmetric profile, the shear
amplitude can be derived as
\begin{eqnarray}
\gamma_M(\theta)&=&\bar{\kappa}(\theta)-\kappa(\theta)
\nonumber\\
&=&4\pi Ga^{-1}\frac{d_A(\chi)d_A(\chi_s-\chi)}{d_A(\chi_s)}
\frac{Mf}{2\pi r_s^2}G(\theta/\theta_{s}),
\label{eqn:shear}
\end{eqnarray}
where $G(x)$ is given by equation (16) in TJ03c. These expressions of
$\kappa_M$ and $\gamma_M$ differ from those given in Bartelmann (1996),
which are derived from an {\em infinite} line-of-size projection of the
NFW profile under the implicit assumption that the profile is valid
(even far) beyond the virial radius.  TJ03b,c carefully verified that
employing the expressions (\ref{eqn:conv}) and (\ref{eqn:shear}) in the
real-space halo model is essential to achieve the model accuracy as well
as the consistency with the Fourier-space halo model well
investigated in the literature (e.g., Seljak 2000). 

Throughout this paper we employ the halo boundary $r_{180}$, not the
virial radius, as stated in \S \ref{modeling}. For this setting, we have
to replace parameter $c$ used in $F(x)$ and $G(x)$ in TJ03b,c with the
ratio of the scale radius to $r_{180}$, $c_{180}$.  The relation between
$c_{180}$ and $c$ is given by $c_{180}=cr_{180}/\rvir$. Note that, even
if we use the virial boundary, the results shown in this paper are
almost unchanged.

Given the convergence and shear profiles for a halo of mass $M$, the
magnification profile is given by
$
\mu_M(\theta)=|(1-\kappa_M(\theta))^2
-\gamma_M^2(\theta)|^{-1}. 
$
This equation implies that $\mu_M$ becomes formally infinity at some
critical radius when the denominator becomes zero.  The radius in the
lens plane is called the critical curve, while the corresponding curve
in the source plane is the caustic curve.

\begin{figure}
  \begin{center}
    \leavevmode\epsfxsize=8.cm \epsfbox{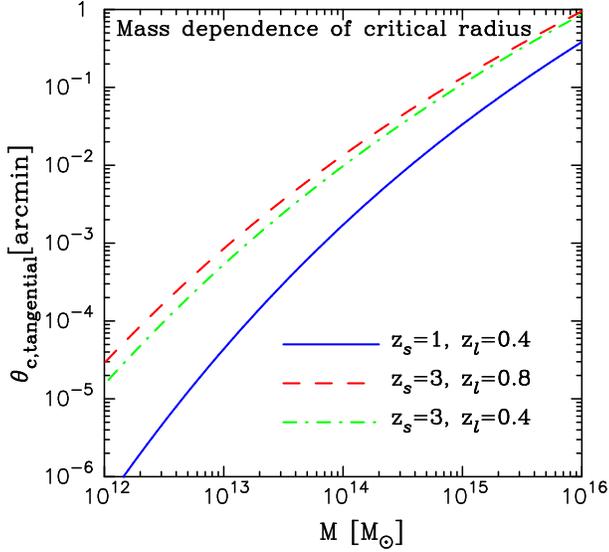}
  \end{center}
\caption{The tangential critical radius of NFW profile against halo
masses.  The solid and dashed curves are the results for lensing halos at
redshifts of $z_l=0.4$ and $0.8$ for source redshifts of $z_s=1$ and
$z_s=3$, respectively.  
Even if
we consider a lower lens redshift, the result
does not largely change, as shown by the dot-dashed curve.  
}
\label{fig:crit}
\end{figure}
Any lensing NFW halo inevitably provides {\em finite} critical radius,
if we have an ideal angular resolution.
This is because the convergence $\kappa_M$ varies from zero to infinity
with changing radius from $\theta_{\rm 180}$ to $0$, while the shear
remains finite over the range (see Figure 2 in TJ03c).  
Figure \ref{fig:crit} plots the tangential critical radius for a lensing
NFW halo against the mass. Note that an NFW profile produces two
critical radii -- an outer curve causes a tangentially deformed image
around it, while the inner one causes a radially deformed image.  The
solid and dashed curves are the results for halos at redshifts $z_l=0.4$
and $0.8$ for source redshifts of $z_s=1.0$ and $3.0$, respectively.
One can see that the critical radius has a strong dependence on halo
masses and is larger for $z_s=3$ than for $z_s=1$ due to the greater
lensing efficiency. These critical curves do not largely change even if
we consider a lower lens redshift than the peak redshift, as shown by
the dot-dashed curve.  Even massive halos with $M\sim 10^{15}M_\odot$ provide
the critical radii of $\simlt 0.\!\!  '1$. 
The scale is below relevant angular scales for the magnification statistics
of our interest. 
However, this does not mean
that the non-linear magnification correction to the correlation function
appears only on scales $\simlt 1'$.  
Rather, modest non-linear magnifications ($\mu\simgt 2$) lead to the
strong impact, since such magnifications have larger cross
section, as will be shown in detail.  

\subsection{Real-space halo model approach}

In the following, we construct the halo model method to compute the
magnification induced correlation function. First, we simply consider
the 2PCF defined as
\begin{equation}
\xi_{\mu}(\theta)
\equiv \kaco{[\mu(\bm{\phi})-1]
[\mu(\bm{\phi}+\bm{\theta})-1]}, 
\label{eqn:auto2pt}
\end{equation}
where $\mu-1$ is the magnification fluctuation field.  This 2PCF is
observable from size fluctuations of distant galaxy images, as discussed
in \S \ref{method}.

From a picture of the halo model, $\xi_\mu(\theta)$ can be expressed as
a sum of correlations between the magnifications fields within a single
halo (1-halo term) and between two different halos (2-halo term):
\begin{equation}
\xi_\mu(\theta)=\xi_{\mu}^{1h}(\theta)+\xi_{\mu}^{2h}(\theta). 
\end{equation}
It is straightforward to extend the real-space halo model developed in
TJ03a,b,c to compute the 1-halo term contribution, which has dominant
contribution on small non-linear scales (see equations (19) and (20) in
TJ03c):
\begin{eqnarray}
\xi_\mu^{1h}(\theta)&=&\int^{\chi_H}_0\!\!d\chi 
\frac{d^2V}{d\chi d\Omega} \int\!\!dM~ n(M)
\nonumber \\
&&\hspace{-3em}\times 
\int^\infty_0\!\! d\phi \int^{2\pi}_0\!\! d\varphi ~ \phi 
\left[\mu_M(\phi)-1\right]
\left[\mu_M(|\bm{\phi}+\bm{\theta}|)-1\right], 
\label{eqn:mu2pt}
\end{eqnarray}
where we have introduced the polar coordinate $d^2\bm{\phi}=\phi d\phi
d\varphi$, $d^2V/d\chi d\Omega=d_A^2(\chi)$ for a flat universe and we
can set the separation vector $\bm{\theta}$ to be along the first axis from
statistical symmetry, thereby
$|\bm{\phi}+\bm{\theta}|=(\phi^2+\theta^2+2\phi\theta\cos\varphi)^{1/2}$.
We have assumed a uniform distribution of sources on the sky and ignored
a probability of multiple images and an increase or decrease in
sampling of the images due to the lensing itself (which is a higher
order correction and can be safely neglected as shown in Hamana 2001).  
The equation above
implies that the 1-halo term contribution is given by summing lensing
contributions due to halos along the line-of-sight weighted with the
halo number density on the light cone. Note that the integration range
of $d^2\bm{\phi}$ is taken as the infinite area, taking into account the
non-local property of the shear field that is non-vanishing at radius
outside the halo boundary.  In practice, setting the upper bound of
$\int\!\!d\phi$ to be three times the projected radius $\theta_{180}$ gives
the same result, to within a few percents.  The validity of the
real-space halo model formulation was carefully investigated in TJ03b,c.

As discussed in \S \ref{lensfield} and shown in Figure \ref{fig:crit},
an NFW profile always provides finite critical curves, where the
magnification formally becomes infinity.  Therefore, to make the halo
model prediction, we introduce a magnification cutoff $\mu_{\rm max}$ in
the calculation -- the integration range of $\int\!\!d^2\bm{\phi}$ is
confined to the region satisfying the condition $\mu_M\le\mu_{\rm max}$
for a given halo.  This procedure is somehow similar to what is done in the
measurement from ray-tracing simulations, where a masking of high
magnification events is employed to avoid a significant statistical
scatter (e.g., see M\'enard et al. 2003; Barber \& Taylor 2002).  Thus
the halo model allows for a fair comparison of the
prediction with the simulation result.
  However, note that the procedure taken 
 ignores the lensing projection effect for simplicity: exactly
speaking, the magnification should be given by the lensing
fields between source and observer, not by those of individual halo.

Similarly,
based on the real-space halo model, the 2-halo term can be expressed as
\begin{eqnarray}
\xi_\mu^{2h}(\theta)&=&\int_0^\infty\!\!d\chi~ \frac{d^2V}{d\chi d\Omega}
\int\!\!dM~ n(M)\int\!\!dM'~ n(M')\nonumber\\
&&\times \int\!\!d^2\bm{\phi}\left[\mu_M(\phi)-1\right]
\int\!\!d^2\bm{\phi'}\left[\mu_{M'}(\phi')-1\right]\nonumber\\
&&\hspace{-4em}\times b(M)b(M')
\int\!\!\frac{ldl}{2\pi}P^L\!\left(k=\frac{l}{d_A}; \chi\right)
J_0(l|\bm{\theta}-\bm{\phi}-\bm{\phi}'|),\nonumber\\
\end{eqnarray}
where $J_0(x)$ is the zero-th order Bessel function and $P^L(k)$ is the
linear mass power spectrum at epoch $\chi$ as given by
$P^L(k;\chi)=D^2(\chi)P(k; t_0)$.  The term including $P^L(k)$ in the
third line on the r.h.s denotes the angular two-point correlation 
function between
different halos of masses $M$ and $M'$, which is derived using Limber's
equation and the flat-sky approximation
(e.g., Blandford et al. 1991; Miralda-Escude 1991; Kaiser 1992).
 We similarly introduce the maximum magnification cutoff in
the 2-halo term calculation as in the 1-halo term.
The equation
above means that we have to perform an 8-dimensional integration
to get the 2-halo term, which is computationally intractable. For this
reason, we employ an approximated way to be valid when angular
separation between the two halos taken is much greater than their 
angular virial radii. This leads to
a simplified expression of the 2-halo term:
\begin{eqnarray}
\xi_\mu^{2h}(\theta)&\approx& \int_0^\infty\!\!d\chi~ 
\frac{d^2V}{d\chi d\Omega}
\nonumber\\
&&\times\left[\int\!\!dM~ b(M)n(M) \int\!\!\phi 
d\phi~ 2\pi(\mu_M(\phi)-1)\right]^2
\nonumber\\
&&\times \int\!\!\frac{ldl}{2\pi}~ P^L\!\left(
k=\frac{l}{d_A(\chi)}; \chi\right)J_0(l\theta). 
\label{eqn:mu2h}
\end{eqnarray}
This allows us to get the 2-halo term by a three-dimensional
integration, because the integrations of the second and third lines on
the r.h.s can be done separately before performing the
$\chi$-integration.  It is worth checking the consistency of the 2-halo
term above with the limiting case $\xi_\mu\approx 4\xi_\kappa$, which is
valid in the weak lensing limit $\mu\approx 1+2\kappa$ for
$\kappa,\gamma\ll 1$.  In this limit, the term contained in the second
line on the r.h.s of equation (\ref{eqn:mu2h}) can be rewritten as
\begin{eqnarray}
\int\!\!dM~ b(M)n(M) \int\!\!d^2\bm{\phi}~ (\mu_M(\phi)-1)&&\nonumber\\
&&\hspace{-15em}\approx \int\!\!dM~ b(M)n(M)\int\!\!d^2\bm{\phi}~ 2\kappa_M(\phi)
\nonumber\\
&&\hspace{-15em}= 2W(\chi)d_A^{-2}(\chi)\int\!\!dM~ b(M)n(M) \frac{M}{\bar{\rho}_0},
\end{eqnarray}
where the second equality is derived from equations (25) and (28) in
TJ03b. Therefore, substituting this result into the 2-halo term
(\ref{eqn:mu2h}) yields
\begin{eqnarray}
\xi_\mu^{2h}(\theta)&\approx& 4\int_0^\infty\!\!d\chi~ 
W^2(\chi)d_A^{-2}(\chi)\nonumber\\
&&\times\left[\int\!\!dM~ b(M)n(M)\frac{M}{\bar{\rho}_0}\right]^2
\nonumber\\
&&\times \int\!\!\frac{ldl}{2\pi}~ P^L\!\left(
k=\frac{l}{d_A(\chi)}; \chi\right)J_0(l\theta)\nonumber\\
&=&4\xi_\kappa^{2h}(\theta). 
\label{eqn:mu2pt2h}
\end{eqnarray}
The 2-halo term (\ref{eqn:mu2h}) is thus reduced to four times the
2-halo term of the convergence 2PCF in the weak lensing limit, as
expected. However, the important point of the 2-halo term
(\ref{eqn:mu2h}) is that it can correctly account for the contribution
of non-linear magnifications ($\mu\simgt 2$) on large scales.

Replacing $P^{2h}(k)$ in equation (\ref{eqn:mu2pt2h}) with a model of
the non-linear mass power spectrum leads to the conventional method for
predicting $4\xi_\kappa(\theta)$ as an estimator of $\xi_\mu(\theta)$,
as employed in the literature (e.g., Dolag \&
Bartelmann 1997; Sanz et al. 1997; Jain et al. 2003).

\section{Asymptotic behavior of magnification statistics 
for high magnifications} \label{asymp}

As stated in the preceding section, we need to introduce a maximum
magnification cutoff in the model prediction to avoid the contribution
from a formally emerged infinite magnification. In practice, strong
lensing events of $\mu \gg 1$, which are identified by
 multiple images or largely deformed image, can be  removed from the
sample of the magnification statistics.
However, without the clear
signatures, it is hard to make the distinct selection and therefore
modest magnification events ($\delta \mu \simgt 1$) are likely included
in the sample for the blind analysis, because the
magnification is not a direct observable. In the following, we clarify 
 how the magnification statistics depend upon
large magnification events ($\mu\gg 1$).

Meanwhile, we restrict our discussion to point-like sources for
simplicity.  High magnifications
$(\mu\gg 1)$ arise from images in the vicinity of the critical curve
that is caused by an intervening mass concentration, such as halos.  For
any finite lens mass distribution, the critical curve must form a closed
non-self-intersecting loop.  Based on the catastrophe theory, it was
shown in Blandford \& Narayan (1986; also see Chapter 6 in Schneider et
al. 1992) that the magnification of an image at a perpendicular distance
$\Delta \theta$ from the (fold) critical curve scales asymptotically as
\begin{equation}
\mu\propto \frac{1}{\Delta \theta}.
\label{eqn:muasy}
\end{equation}
This argument holds independently of details of the lensing mass
distribution, although the proportional coefficient does depend on the
mass distribution.

To keep generality of our discussion, let us consider a correlation
function between the magnification field and some cosmic fields
expressed as
\begin{eqnarray}
\xi(\theta; p)=\skaco{[\mu(\bm{\theta}_1)]^p
f(\bm{\theta}_2)}_{|\bm{\theta}_1-\bm{\theta}_2|=\theta},
\label{eqn:mustat}
\end{eqnarray}
where $p$ is an arbitrary number.  The field $f$ is allowed to be
constituted from any cosmic fields which are correlated with the
magnification field. Therefore, the following
argument holds for high-order moments beyond the two-point correlations if
one takes products of the cosmic fields for $f$,
e.g. $f=\delta(\bm{\theta}_2) \delta(\bm{\theta}_3)$.

Suppose that an intervening halo provides the critical curve in the lens
plane for a given source redshift, as this is the case for an NFW halo (see
Figure \ref{fig:crit}).  Then, let us consider how high magnifications
in the vicinity of this critical curve contribute to the magnification
statistics.  By introducing an upper bound on the magnification fields
as $\mu\le \mu_{\rm max}$, we can address how the magnification
statistics depend on the cutoff $\mu_{\rm max}$ and what is the
asymptotic behavior for the limiting case $\mu_{\rm max}\rightarrow
\infty$.  From equation (\ref{eqn:muasy}), the cutoff $\mu_{\rm max}$
corresponds to a lower limit on the perpendicular distance from the
critical curve, say $\Delta\theta\ge \epsilon$ ($\epsilon\rightarrow 0$
corresponds to $\mu_{\rm max}\rightarrow \infty $). As can be seen from
equations (\ref{eqn:mu2pt}) and (\ref{eqn:mu2h}), a picture of the halo
model leads us to compute the high magnification contribution to the
magnification-induced correlation like equation
(\ref{eqn:mustat}) by the integration
\begin{equation}
\xi(\theta; p, \mu_{\rm max})
\sim \int_{\partial S_c,\mu\le\mu_{\rm max}}\!d^2\bm{s}~
 \frac{1}{|\Delta s|^p}
f(|\bm{s}+\bm{\theta}|),
\label{eqn:muint}
\end{equation}
where $\Delta s$ is the perpendicular distance from the critical curve
and the integration range is confined to the area $\partial S_c$ subject
to the condition $\mu\le \mu_{\rm max}$.  The integration
(\ref{eqn:muint}) results in one-dimensional integration for high
magnifications around the critical curve, in analogy with equation
(5.16) in Blandford \& Narayan (1986) to derive the asymptotic, integral
cross section for the strong lensing events that produce multiple
images\footnote{The discussion in this paper as well as in Blandford \&
Narayan (1986) employs the assumption that asymptotic dependence of the
magnification statistics on high magnifications is mainly due to images
around the fold caustics and therefore ignores the contribution from the
cusp caustic. This is likely to be a good approximation as shown in Mao
(1992).}.  Hence, the leading order contribution of $\mu_{\rm max}$ can
be expressed as
\begin{eqnarray}
\xi(\theta; p,\mu_{\rm max})
&\sim&
\left\{
\begin{array}{ll}
1/\mu_{\rm max}^{1-p}, & p<1\\
\ln(\mu_{\rm max}), & p=1\\
\mu_{\rm max}^{p-1}, & p>1.
\end{array}
\right.
\label{eqn:muasymp}
\end{eqnarray}
where we have assumed that variation in the field $f$ does not largely
change for the relevant integration range. 
The equation above leads to
an intriguing consequence: the amplitude of the magnification
correlation is finite for $p<1$ for the limiting case $\mu_{\rm
max}\rightarrow \infty$, while it diverges for $p\ge 1$. 
Thus, the statistics with $p<1$ is practically advantageous in that it
is insensitive to the uncertainty of which magnification cutoff should be
imposed for a given sample. 
Furthermore,
the asymptotic behavior does not explicitly depend on the separation
angle $\theta$ and therefore it holds even for large $\theta$. This
means that the divergence $\xi\rightarrow \infty$ for $p\ge 1$ formally 
occurs even on degree scales, which is opposed to a naive expectation that the
weak lensing approximation is safely valid on these scales.  It is
also worth stressing that this behavior is expected to hold for any
lensing mass distribution once the critical curve appears, although the
proportionality coefficient of $\xi$ should depend on details of the
mass distribution.  These will be quantitatively
tested by the halo model prediction as well as by the ray-tracing
simulation.

In reality, a finite source size imposes a maximum cutoff on the
observed magnification and thus the infinite magnification does not
happen, even if a source sits on the caustic curve in the source plane
(see Chapters 6 and 7 in Schneider et al. 1992; Peacock 1982; Blandford
\& Narayan 1996).  For example, if the source is a circular disk with
radius $r_S$ and uniform surface brightness, the maximum magnification
is given by $\mu_{\rm max}\propto 1/\sqrt{r_S}$.  Therefore, to develop
an accurate model prediction requires a knowledge of unlensed source
properties such as the size and the surface brightness distribution
in addition to modeling the lensing mass distribution. In
particular, this could be crucial if one considers the magnification
statistics (\ref{eqn:mustat}) with $p\ge 1$, since it is 
sensitive to large magnification events.

\section{Results}
\label{results}

\subsection{Ray-tracing simulations}
To test the analytic method developed in \S \ref{model}, we employ the
ray-tracing simulations.  We will use the simulation for the current
concordance \LCDM model with $\Omega_{\rm m0}=0.3$,
$\Omega_{\lambda0}=0.7$, $\Omega_{\rm b0}=0.04$, $h=0.7$ and $\sigma_8=0.9$
(M\'enard et al. 2003; Hamana et al. in preparation; TJ03c).  The $N$-body
simulations were carried out by the Virgo Consortium \footnote{ see {\tt
http://star-www.dur.ac.uk/\~{}frazerp/virgo/virgo.html} for the details}
(see also Yoshida, Sheth \& Diaferio 2001), 
and were run using the
particle-particle/particle-mesh (P$^3$M) code with a force softening
length of $l_{\rm soft}\sim 30h^{-1}$kpc. 
The initial matter power spectrum was computed using CMBFAST (Seljak \&
Zaldarriaga 1996).
For the analytic model, we approximate the
initial condition to use the CDM transfer function given by 
Bardeen et al. (1986) with the shape
parameter in Sugiyama (1995) for simplicity.  The $N$-body simulation
employs $512^3$ CDM particles in a cubic box of $479h^{-1}$Mpc on a
side, and the particle mass of the simulation is $m_{\rm part}=6.8\times
10^{10}h^{-1}M_\odot$.

The multiple-lens plane algorithm to simulate the lensing maps from the
$N$-body simulations is detailed in Jain et al. (2000) and Hamana \&
Mellier (2001).  We will use the output data for source redshifts of
$z_s=1$ and 3 for the following analysis.  The simulated map is given on
$1024^2$ grids of a size of $\theta_{\rm grid}=0.\!  \! '2$; the area is
$\Omega_s=11.7$ degree$^2$.  The angular resolution that is unlikely
affected by the discreteness of the $N$-body simulation is around $1'$
(M\'enard et al. 2003; TJ03c).

\subsection{Probability distribution function
 of the magnification: the angular resolution of the simulations}
\begin{figure}
  \begin{center}
    \leavevmode\epsfxsize=8.cm \epsfbox{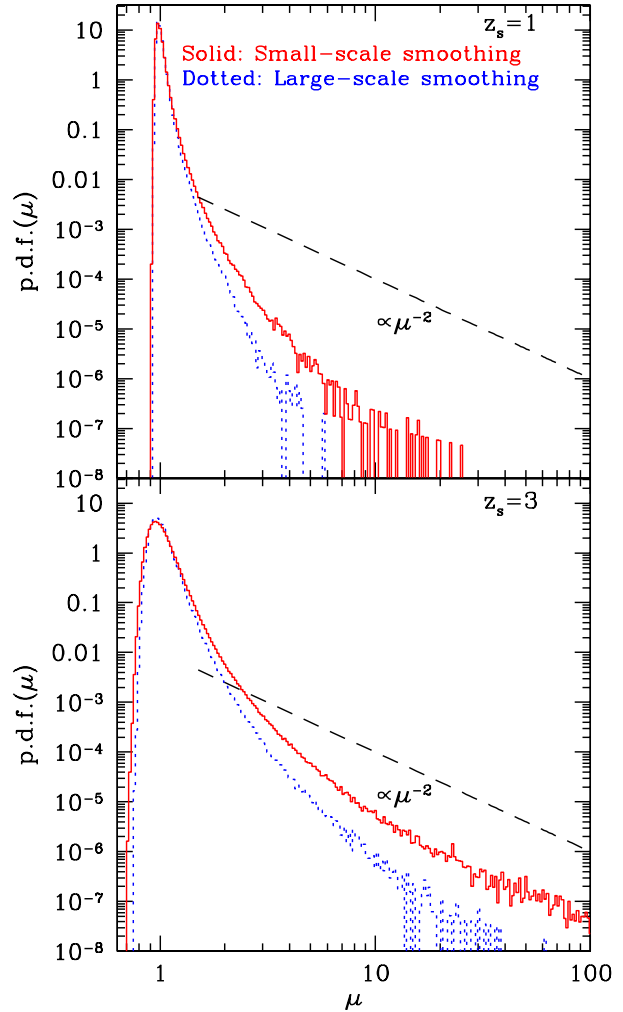}
  \end{center}
\caption{The probability density function (PDF) of the magnification
measured from the ray-tracing simulations for $z_s=1$ and $3$.  To
clarify the angular resolution, we show the two results of using
different smoothing scales, which were used to avoid the artificial
discrete effect of the $N$-body simulations.  The comparison manifests
that high magnification events are sensitive to smaller structures that
are relevant for the smoothing.  The dashed line shows an
asymptotic behavior of the PDF for high magnification events, ${\rm PDF}
\propto \mu^{-2}$, as theoretically expected (see text in more detail).
} \label{fig:pdfzs}
\end{figure}
Universal properties of the critical curve, as demonstrated in \S
\ref{asymp}, lead to an asymptotic behavior of the probability
distribution function (PDF) of large magnification events, irrespective
of details of the lensing mass distribution: ${\rm PDF}(\mu)\propto
\mu^{-2}$ (Peacock 1982; Vierti \& Ostriker 1983;
Blandford \& Narayan 1986; Schneider 1987; 
Hamana, Martel \& Futamase 2000, 
and also see Chapters
6, 11 and 12 of Schneider et al. 1992). Note that the PDF is defined in
the image plane, while the PDF becomes $\mu^{-3}$ if one
defines it from the sample of sources 
in the source plane.
We use this property to
investigate the angular resolution of the ray-tracing simulation.

Figures \ref{fig:pdfzs} shows the magnification PDF measured in the
simulations for source redshifts of $z_s=1$ and $3$, respectively. To
compute the PDF, we accumulate the counts in a given bin of $\mu$ from
36 realizations and then normalize the PDF amplitude to satisfy
$\int\!\!d\mu~ {\rm PDF}(\mu)=1$ over the range of $\mu$ measured.  The
PDF has a skewed distribution: most events lie in demagnification of
$\mu<1$ and rare events have high magnifications with $\mu\gg 1$ having
a long tail. These reflect an asymmetric mass distribution in the
large-scale structure as expected from the CDM scenario -- the
underdense region can be seen preferentially in the void region with a
typical size $\sim 10$Mpc, while the highly non-linear structures appear
in dark matter halos on scales $\simlt $Mpc.  As can be seen, the
simulation of $z_s=3$ displays more pronounced evidence of asymptotic
dependence ${\rm PDF}(\mu)\propto \mu^{-2}$ for high magnifications 
($\mu\simgt 10$) than
the result for $z_s=1$.

To more explicitly clarify the resolution issue of the simulations, we
show the two results of the different smoothing scales that were used in
making the projected density field to suppress the discreteness effect
of the $N$-body simulations (see M\'enard et al. 2003 and Hamana et
al. 2003 for more details).  They are named as `small-scale smoothing'
(solid curve) and `large-scale smoothing' (dotted curve), respectively.
The former is expected to have the angular resolution around $1'$ as
discussed in M\'enard et al. (2003) and in TJ03c, while the latter
employs a smoothing scale two times larger than the former.  The effect
of the large-scale smoothing is that it more smoothes out smaller scale
structures of the mass distribution that are resolved by the small-scale
smoothing simulation. The comparison manifests that occurrence of high
magnification events ($\mu>1$) is very sensitive to the small-scale
structures.  For this reason, we will employ the small-smoothing
simulation in the following, since our interest is to clarify the
non-linear magnification effect
on the magnification statistics.  
This result also implies that simulations
with higher resolution could further alter the PDF shape especially at
high magnification tail.

\subsection{The two-point correlation function of lensing magnification}
\begin{figure}
  \begin{center}
    \leavevmode\epsfxsize=8.cm \epsfbox{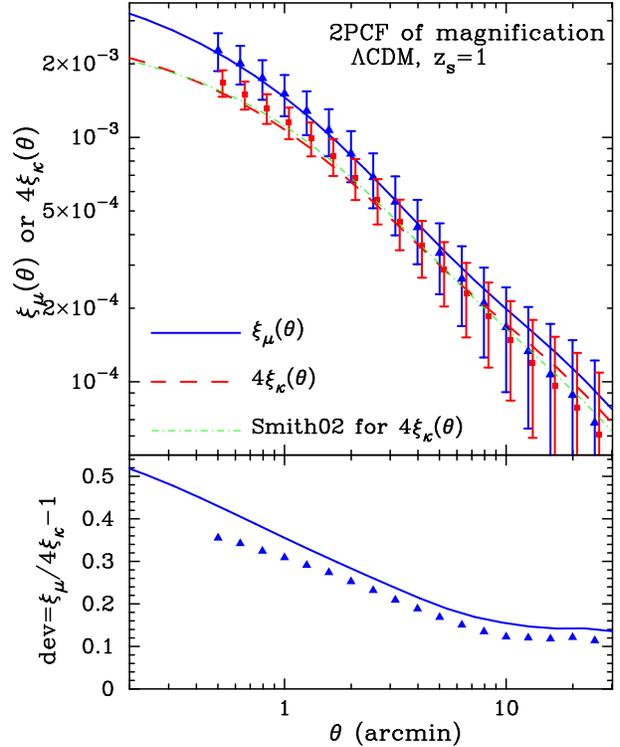}
  \end{center}
\caption{The two-point correlation function of the magnification field
against the separation angle for the \LCDM model and source redshift
$z_s=1$. 
The solid curve shows the halo model prediction, while the triangle
symbol is the simulation result. In most part of this paper, 
the maximum
magnification cutoff $\mu_{\rm max}=8$ is employed 
(see text for the details).
The error bar denotes the sample
variance for a simulated area of $11.7$ degree$^2$.  For comparison, the
weak lensing predictions, leading to $\xi_\mu\approx
4\skaco{\kappa\kappa}$, are shown by the halo model (dashed curve), the
simulation (square symbol) and the fitting formula of Smith et
al. (2003; dot-dashed curve). Note that the simulation result is
slightly shifted in the horizontal direction for illustrative purpose.
The lower panel explicitly 
plots the contribution of non-linear magnifications,
$\xi_\mu/4\xi_\kappa-1$, for the halo model prediction and the
simulation result.  } \label{fig:compzs1}
\end{figure}
\begin{figure}
  \begin{center}
    \leavevmode\epsfxsize=8.cm \epsfbox{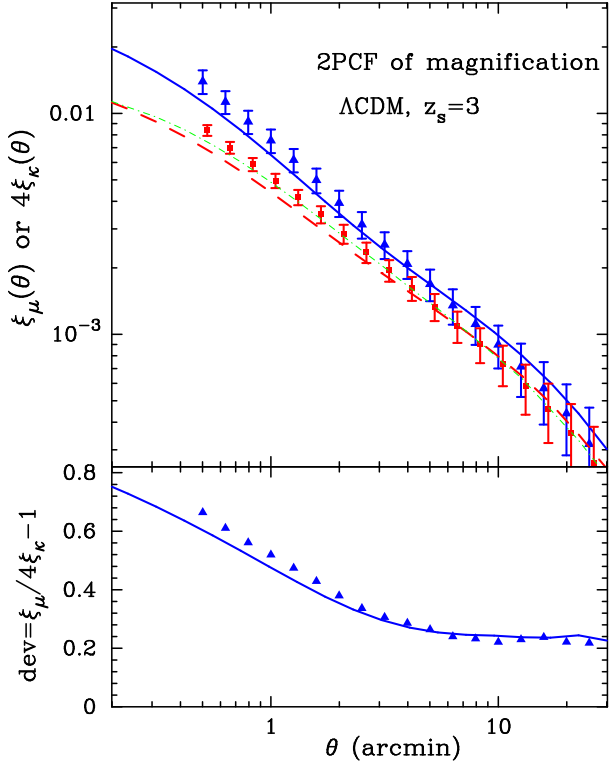}
  \end{center}
\caption{As in the previous figure, but for $z_s=3$.  For higher source
redshifts, the non-linear magnification effect becomes more significant as
expected.  
}
\label{fig:compzs3}
\end{figure}

We now turn to investigation of the magnification 2PCF,
$\xi_\mu\equiv\skaco{\delta\mu\cdot\delta\mu}$, as it is possible to
observe from size fluctuations on distant galaxy images (see \S
\ref{method}).  Figure
\ref{fig:compzs1} shows the comparison of the halo model prediction
(solid curve) with the measurement from simulations (triangle symbol)
for source redshift $z_s=1$.  Note that the error bar in each bin denotes
the sample variance for a simulated area of $11.7$ degree$^2$, which
is computed from $36$ realizations,
and  the errors in neighboring bins are highly correlated. 
In this and following results, we mainly employ the maximum
magnification cutoff $\mu_{\rm max}=8$ in the halo model prediction as
well as in the simulation result.  If we ignore the shear contribution
to the magnification (\ref{eqn:mag}), this cutoff corresponds to
$\mu\approx 1+2\kappa= 2.3$ for the weak lensing approximation. 
 The cutoff value is
chosen so that strong lensing events are removed from the analysis,
because such events likely have greater magnification $\mu\simgt 10$
(private communication with M. Oguri). Figure \ref{fig:pdfzs} shows that
 this cutoff leads us to exclude the events in a high magnification tail
of the PDF.  

Figure \ref{fig:compzs1} shows that
 the halo model prediction well matches the simulation
result.  The 1-halo term provides dominant contribution to the total
power on small scales $\simlt 3'$, while the 2-halo term eventually
captures the larger scale signal (see Figure A1 of TJ03c).  It is worth
noting that the shear field in $\mu$ (see equation (\ref{eqn:mag}))
contributes to the 2PCF amplitude by $\sim 10\%$ over the scales
considered.  To make clear the importance of the non-linear
magnification contribution ($\delta \mu\simgt 1$), the dashed curve and the
square symbol are the halo model prediction and the simulation result
 for the weak lensing approximation $\xi_{\mu} \approx
4\skaco{\kappa\kappa}=4\xi_\kappa$.  For this case, $4\xi_\kappa$ can be
also computed from the fitting formula of the non-linear mass power
spectrum recently proposed by Smith et al.  (2003; hereafter Smith03),
which demonstrates another test of the accuracy of the halo model as
well as of the simulation.  

The lower panel explicitly plots 
the relative difference,
$\xi_\mu/4\xi_\kappa-1$.  The
simulation result is computed from the mean values of $\xi_\mu$ and
$4\xi_\kappa$, and we do not plot the large error bar for illustrative
purpose. 
The correction of the non-linear magnification
 amounts to $\simgt 30\%$ at $\theta\simlt 2'$, and the
non-negligible contribution of $\sim 10\%$ still remains even on large
scales $\simgt 10'$.  
Our model of the 2-halo term (\ref{eqn:mu2h}) correctly captures
the non-linear effect seen in the simulation on the large scales. 
 This large-scale correction
is somewhat surprising, since it is naively
expected that the weak lensing approximation is valid at these scales.  
M\'enard et
al. (2003) also showed that the non-linear correction on the large scales
can be fairly explained taking into account the higher-order terms
$O(\kappa^2)$ in the Taylor expansion of $\mu$, though the method
ceases to be accurate on small scales $\simlt 5'$.  
One advantage of the halo model is that it allows us to explicitly
introduce the maximum magnification cutoff in the calculation,
which allows a fair comparison with the simulation and probably  with 
the actual observation. In other words,
the results shown depend upon the cutoff value employed. 
If we use the cutoff values of $\mu_{\rm
max}=2$ and $100$, the deviation, $\xi_\mu/4\xi_\kappa-1$, becomes 
$\sim 20\%$ and $\sim 40\%$ at $\theta=1'$, 
respectively (also see Figure \ref{fig:mumax}). 

Figure \ref{fig:compzs3} shows the result for $z_s=3$, as in the
previous figure. It is clear that the non-linear magnification 
contribution
leads to significant enhancement in the amplitude of the magnification
correlation relative to the weak lensing prediction.  
The enhancement is $\simgt 40\%$ at
$\theta\simlt 2'$.  The comparison with the previous figure manifests
that sources at higher redshifts are more affected by the non-linear
magnification, as the sources have more chance to encounter intervening
halos. It is worth noting that, if we do not apply any masking of high
magnification events in the simulation, 
the statistical error in each bin becomes very large, which indicates
the presence of events with $\mu\gg 1$ in some realizations as shown 
 in Figure
\ref{fig:pdfzs}.

\subsection{A quantitative test of the asymptotic behavior 
of the magnification statistics}
\begin{figure}
  \begin{center}
    \leavevmode\epsfxsize=8.5cm \epsfbox{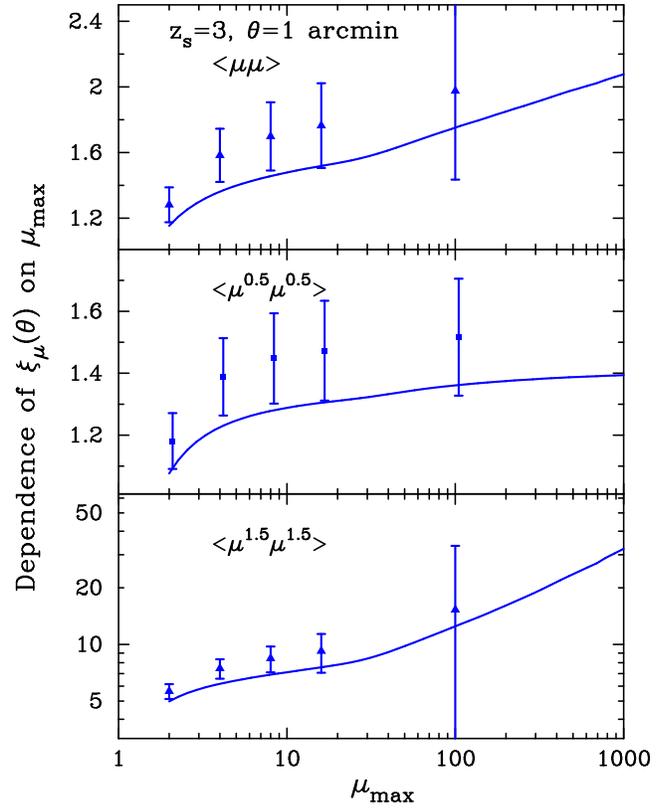}
  \end{center}
\caption{Dependences of the magnification 2PCF amplitude 
on the maximum magnification cutoff
used in the evaluations.  For the magnification correlation
parameterized as $\skaco{\delta\mu^p\cdot \delta\mu^p}$, the three panels
show the results for $p=1$, $0.5$ and $1.5$, 
respectively.  The source redshift  $z_s=3$ and the separation angle 
$\theta=1'$ are considered. All the curves are
normalized by the predictions derived from the weak lensing
 approximation $(\mu\approx 1+2\kappa)$. 
As discussed in \S \ref{asymp}, it is
expected that the 2PCF amplitude diverges for $p\ge 1$ for the limiting
case $\mu_{\rm max}\rightarrow \infty$, while
it remains finite for $p<1$ (see equation (\ref{eqn:muasymp})).  This is
verified by the halo model as well as by the simulation.  }
\label{fig:mumax}
\end{figure}
In what follows we quantitatively  test the asymptotic
dependence of the magnification statistics on large magnifications
($\mu\gg 1$), as derived in \S \ref{asymp}.  For this purpose, we
consider three cases of $p=0.5$, $1$ and $1.5$ for the magnification
2PCF parameterized as $\xi=\skaco{\delta\!\mu^p\cdot\delta\!\mu^p}$,
with source redshift $z_s=3$.  Figure \ref{fig:mumax} shows how the 2PCF
amplitude depends on the magnification cutoff $\mu_{\rm max}$ used in
the halo model predictions and the simulation result. 
The separation angle of $\theta=1'$ is considered and
the curves are normalized by the predictions from
 the weak lensing approximation. The scale $\theta=1' $ is chosen based
on the fact that the scale is in the non-linear regime and
unlikely affected by the angular resolution of the simulation (TJ03c).
First, one can see that even the most conservative choice of 
$\mu_{\rm max}=2$ leads to decent difference between the 
correct treatment and
the weak lensing approximation.
The
consequence derived in \S \ref{asymp} 
is that the 2PCF amplitudes for $p=0.5$,
$1.0$ and $1.5$ have the dependences on $\mu_{\rm max}$ given as
$\xi\propto \mu_{\rm max}^{-0.5}$, $\ln(\mu_{\rm max})$ and $\mu_{\rm
max}^{0.5}$ for $\mu_{\rm max}\gg 1$, respectively.  It is obvious that
this consequence is verified by the halo model prediction as well as by the
simulation result for $\mu_{\rm max}\simgt 10$. Although the halo model
results for $\alpha=1.0$ and $1.5$ display a bend at $\mu_{\rm
max}\approx 30$, we found that 
this is due to high magnifications between the radial
and tangential critical curves in NFW halos.  Most importantly, the 2PCF
amplitude for $p=0.5$ has a well converged value for $\mu_{\rm
max}\simgt 5$: the amplitude changes by less than $10\%$ over $\mu_{\rm
max}=[10,1000]$. 
Therefore, the statistics with $p<1$ have a practically
great advantage because it is little affected by the uncertainty in
specifying the maximum magnification cutoff in the analysis.

Finally, we note that the results shown above are unchanged even if we
consider the cross-correlation $\skaco{\delta\!\mu^p\cdot \delta^{2D}}$,
where $\delta^{2D}$ is the projected density fluctuation field (e.g.,
see equation (\ref{eqn:delg})),  because the asymptotic behavior is
determined by the power of $\mu$ entering into the general correlation
function $\skaco{\mu^p f}$, as derived in \S \ref{asymp}.

\subsection{Application to  QSO-galaxy cross correlation}
\label{qso}
\begin{figure*}
  \begin{center}
    \leavevmode\epsfxsize=16.5cm
    \epsfbox{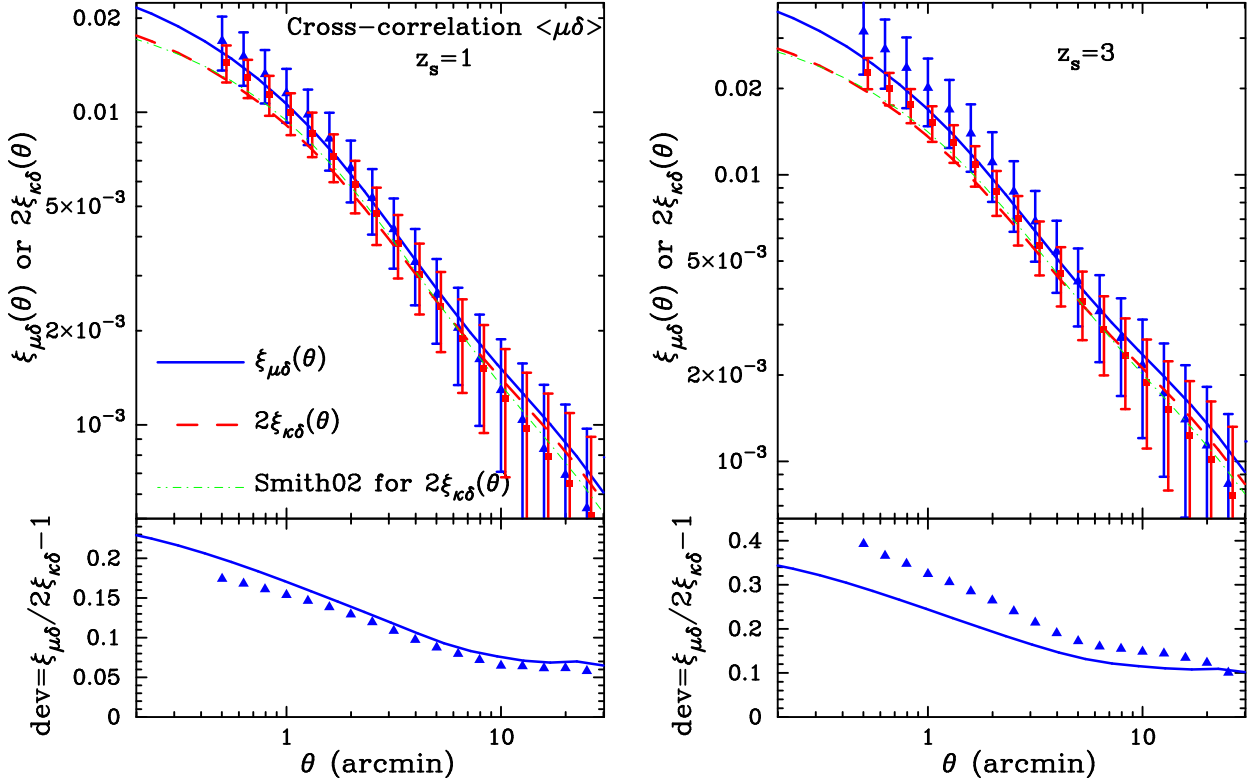}
  \end{center}
\caption{The cross correlation function between the magnification field
and the projected density fluctuation field for $z_s=1$ (left panel) and
 3 (right panel), as in Figure
\ref{fig:compzs1}. To get the projected density field, we assume the
redshift selection function given by equation (\ref{eqn:select}).
}\label{fig:md2ptzs1}
\end{figure*}
In this subsection, we consider an application of the halo model to the
QSO-galaxy cross correlation.  The angular fluctuation field of galaxies
on the sky is a projection of the 3D galaxy fluctuation field $\delta^g$
along the line-of-sight, weighted with the redshift selection function
$f_g(z)$ of the galaxy sample:
\begin{equation}
\delta n^g(\bar{\theta})=\int^{\infty}_0\!\!dz~ f_g(z)
~ \delta^g(d_A(z)\bm{\theta},z),
\label{eqn:delg}
\end{equation}
where $f_g(z)$ is normalized as $\int^\infty_0\! dz
f_g(z)=1$. Throughout this paper, we employ
\begin{equation}
f_g(z)dz=\frac{\beta z^2}{z_0^3\Gamma(3/\beta)}\exp\left[-\left(
\frac{z}{z_0}\right)^\beta\right]dz,
\label{eqn:select}
\end{equation}
with $\beta=1.5$ and $z_0=0.3$. This model leads to the mean redshift 
as $z_{\rm mean}=\int\!\!dz~ zf_g(z)=0.45 $ and
roughly reproduces the actual
distribution of galaxies in the redshift galaxy
catalog (e.g., Dodelson et al. 2002). 

However, the galaxy fluctuation field $\delta^g$
is not straightforward to model, since the galaxy formation is affected by
complex astrophysical processes in addition to the gravitational effect.
Recently, Jain et al. (2003) developed a sophisticated description of
the magnification correlations based on the halo model as well
as the semi-analytic galaxy formation model.  
In particular, it was shown that it
is crucial to account for a realistic model to describe how galaxies populate
their parent halo, the so-called halo occupation number, to make
the accurate model predictions on arcminute angular scales.  The halo
occupation number strongly depends on types of
galaxies such red or blue galaxies.  We here address how 
the non-linear magnification further modifies the model prediction. 

Before going to this study, we consider the cross-correlation between the
magnification field and the dark matter distribution, which corresponds
to an unrealistic case that the galaxy distribution exactly traces the
underlying mass distribution; $\delta_{\rm g}=\delta$.  
This investigation is aimed at clarifying how
the non-linear magnification effect remains after inclusion of a
realistic model of the galaxy clustering, from the comparison of the results
with and without the galaxy bias model.  
In
addition, in this case we can compare the model prediction with the
simulation result that is computed from the same
$N$-body simulation we have used.
Extending the method presented in \S \ref{model} leads to the 1-halo
term contribution to the cross-correlation:
\begin{eqnarray}
\xi^{1h}_{\mu\delta}(\theta)&=&\int^{\chi_H}_0\!\!d\chi 
\frac{d^2V}{d\chi d\Omega} f_g(z)\frac{dz}{d\chi}
\int\!\!dM~ n(M)\frac{M}{\bar{\rho}_0}
\nonumber \\
&&\hspace{-3em}\times 
\int^\infty_0\!\! d\phi \int^{2\pi}_0\!\! d\varphi ~ \phi 
\left[\mu_M^{2.5s-1}(|\bm{\phi}+\bm{\theta}|)
-1\right]\Sigma_M(\phi),
\label{eqn:mudelta}
\end{eqnarray}
where $\Sigma_M(x)$ is the normalized projected density of the NFW
profile given by equation (26) in TJ03b. Similarly, one can derive the
2-halo term of $\xi_{\mu\delta}$, as done by equation (\ref{eqn:mu2h}).

In Figure \ref{fig:md2ptzs1} we shows the results for
source redshifts of $z_s=1$ (left panel) and $3$ (right panel), as in Figures
\ref{fig:compzs1} and \ref{fig:compzs3}. Note that both the results
employ the same redshift selection function (\ref{eqn:select}) to obtain
the projected density field. We simply assumed a special case of $s=4/5$
for the magnitude slope for the unlensed QSO number count.  From the
comparison with Figures \ref{fig:compzs1} and \ref{fig:compzs3}, it is
clear that the non-linear magnification contribution is weakened, due to the
single power of $\mu$ entering into the two-point correlation compared
to the magnification 2PCF.  Nevertheless, it should be stressed that the
non-linear correction has significant contributions of $\simgt 15\%$ and
$\simgt 25\%$ at $\theta\simlt 1'$ for $z_s=1$ and 3, respectively.

Next, we consider a model of the QSO-galaxy correlation that takes into
account both the galaxy bias and the non-linear magnification effect. 
To do this, we use the halo occupation number $\skaco{N_{\rm g}(M)}$ to
describe how many galaxies populate their parent halo of a given
mass $M$, in an average sense (e.g., Seljak 2000;
 Guzik \& Seljak 2002; Jain et al. 2003; TJ03b; Cooray \& Sheth 2002).
Simply replacing $M/\bar{\rho}_0$ in equation (\ref{eqn:mudelta}) with 
$\skaco{N_{\rm g}(M)}/\bar{n}_{\rm gal}$ leads to
the 1-halo term of the QSO-galaxy cross correlation:
\begin{eqnarray}
\xi^{1h}_{\mu {\rm g}}(\theta)&=&
\int^{\chi_H}_0\!\!d\chi 
\frac{d^2V}{d\chi d\Omega} f_g(z)\frac{dz}{d\chi}
\int\!\!dM~ n(M)\frac{\skaco{N_g(M)}}{\bar{n}_{\rm gal}}
\nonumber \\
&&\hspace{-3em}\times 
\int^\infty_0\!\! d\phi \int^{2\pi}_0\!\! d\varphi ~ \phi 
\left[\mu_M^{2.5s-1}(\phi)
-1\right]
\Sigma_M(|\bm{\phi}+\bm{\theta}|),
\label{eqn:gbias}
\end{eqnarray}
where $\bar{n}_{\rm gal}$ is the average number density at epoch $z$
defined as $\bar{n}_{\rm gal}=\int\!\! dM~ n(M)\skaco{N_{\rm g}(M)}$.
The cross-correlation thus depends on the first moment of $\skaco{N_{\rm
g}(M)}$ \footnote{This is also the case for galaxy-galaxy lensing as
shown in Guzik \& Seljak (2002)}. Note that, on the other hand, 
 the
two-point correlation of galaxies depends on the second moment, and
it contains somehow uncertainty in modeling the sub-Poissonian process
in the regime of $\skaco{N_{\rm g}(M)}<1$.  

As stressed in Guzik \& Seljak (2002) and Jain et al. (2003),
it is probably accurate to assume that one of the $\skaco{N_{\rm g}}$
galaxies in a halo sits at the halo center and this has
decent impact on the model predictions. On the other hand,
we assume that the other $(\skaco{N_{\rm g}}-1)$ galaxies 
follow the dark matter distribution within the halo. 
Following the method of Jain et al. (2003), the part of the integrand
function in the 1-halo term (\ref{eqn:gbias}) can be expressed as
\begin{eqnarray}
&&\hspace{-2em}\skaco{N_g(M)}
\int^\infty_0\!\! d\phi \int^{2\pi}_0\!\! d\varphi ~ \phi 
\left[\mu_M^{2.5s-1}(\phi)-1\right]
\Sigma_M(|\bm{\phi}+\bm{\theta}|)\nonumber\\
&=&\chi^{-2}\left[\mu_M^{2.5s-1}(\theta)-1\right]
\nonumber\\
&&+\int^\infty_0\!\! d\phi \int^{2\pi}_0\!\! d\varphi ~ \phi 
\left[\mu_M^{2.5s-1}(\phi)-1\right]
\Sigma_M(|\bm{\phi}+\bm{\theta}|)\nonumber\\
&&\times\left[\skaco{N_g(M)}-1\right],
\end{eqnarray}
for $\skaco{N_g(M)}\ge 1$ and
\begin{eqnarray}
&=&\left[\mu_M^{2.5s-1}(\theta)-1\right]
\skaco{N_g(M)},
\end{eqnarray}
for  $\skaco{N_g(M)}< 1$.  Substituting these equations into equation 
(\ref{eqn:gbias}) leads to the halo model prediction for the QSO-galaxy
cross-correlation. 

To complete the model prediction, we need an adequately accurate model
of the halo occupation number $\skaco{N_{\rm g}(M)}$. 
We employed the model in
Jain et al. (2003), which was derived from the GIF $N$-body simulations,
coupled to a semi-analytic galaxy formation model (Kauffmann et
al. 1999). The simulation result of $\skaco{N_{\rm g}(M)}$ is well
fitted by the functional form,
\begin{equation}
\skaco{N_{g}(M)}=\left(\frac{M}{M_0}\right)^\alpha+
A\exp\left[-A_0\left(\log_{10}(M)-M_B\right)^2\right].
\label{eqn:occup}
\end{equation}
The parameter values are taken from Table 1 labeled as `$z=0.06$' in Jain et
al. (2003), which reproduces the measurements for total (blue plus red)
galaxies at $z=0.06$ in the GIF simulations.
 In the following, we employ a lower
mass cutoff of $M\ge 10^{11}~ h^{-1}M_\odot$ and ignore the redshift
evolution of $\skaco{N_{\rm g}(M)}$ for simplicity. This model leads to
the galaxy bias parameter $\bar{b}_{\rm gal}=(1/\bar{n}_{\rm
gal})\int\!dM n(M)b(M)\skaco{N_{\rm g}(M)}= 1.2$ at $z=0$ in the
large-scale limit, thus
reflecting the fact that the modeled galaxies are biased objects relative
to the dark matter distribution.

\begin{figure*}
  \begin{center}
    \leavevmode\epsfxsize=16.5cm 
    \epsfbox{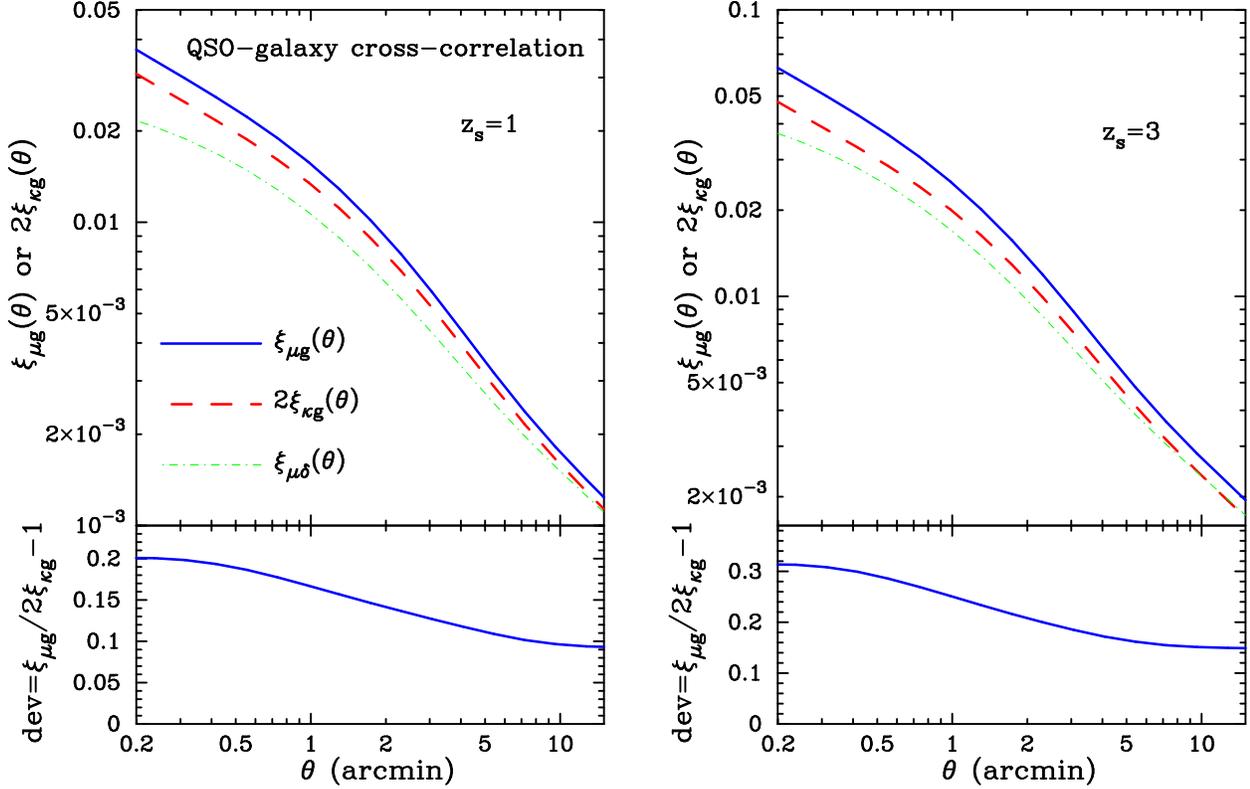}
  \end{center}
\caption{The halo model predictions for the angular cross-correlation
between the magnification and the foreground galaxy distribution, where
the galaxy clustering is modeled by the halo occupation number
(\ref{eqn:occup}).  For comparison, the dot-dashed
curve is the result for the cross-correlation between the magnification
and the projected dark matter distribution as shown in the previous
 figure.  
The non-linear magnification correction remains to enhance the
cross-correlation amplitude relative to the weak lensing approximation,
 even if the realistic model of the galaxy clustering is included.
}\label{fig:gbias}
\end{figure*}
Figure \ref{fig:gbias} shows the model predictions for the QSO-galaxy
cross-correlation, as in the previous figure.  For comparison, the
dot-dashed curve is the result for the cross-correlation between the
magnification and the projected dark matter distribution in the previous
figure. The comparison of the solid and dot-dashed curves manifests
that the realistic model of galaxy bias
largely modifies the cross-correlation, and the
galaxy bias cannot be described by a simple linear bias on the small
angular scales (Jain et al. 2003).  It is also clear that the non-linear
magnification correction is $10-25\%$ on arcminute
scales.
 Therefore, this result implies that an inclusion of
the non-linear effect will be necessary to make an unbiased
interpretation of the precise measurement expected from forthcoming
massive surveys such as the CFHT Legacy Survey and the SDSS. 

\subsection{Sensitivity of the magnification statistics to the halo
  profile properties}
\begin{figure*}
  \begin{center}
    \leavevmode\epsfxsize=16.5cm 
    \epsfbox{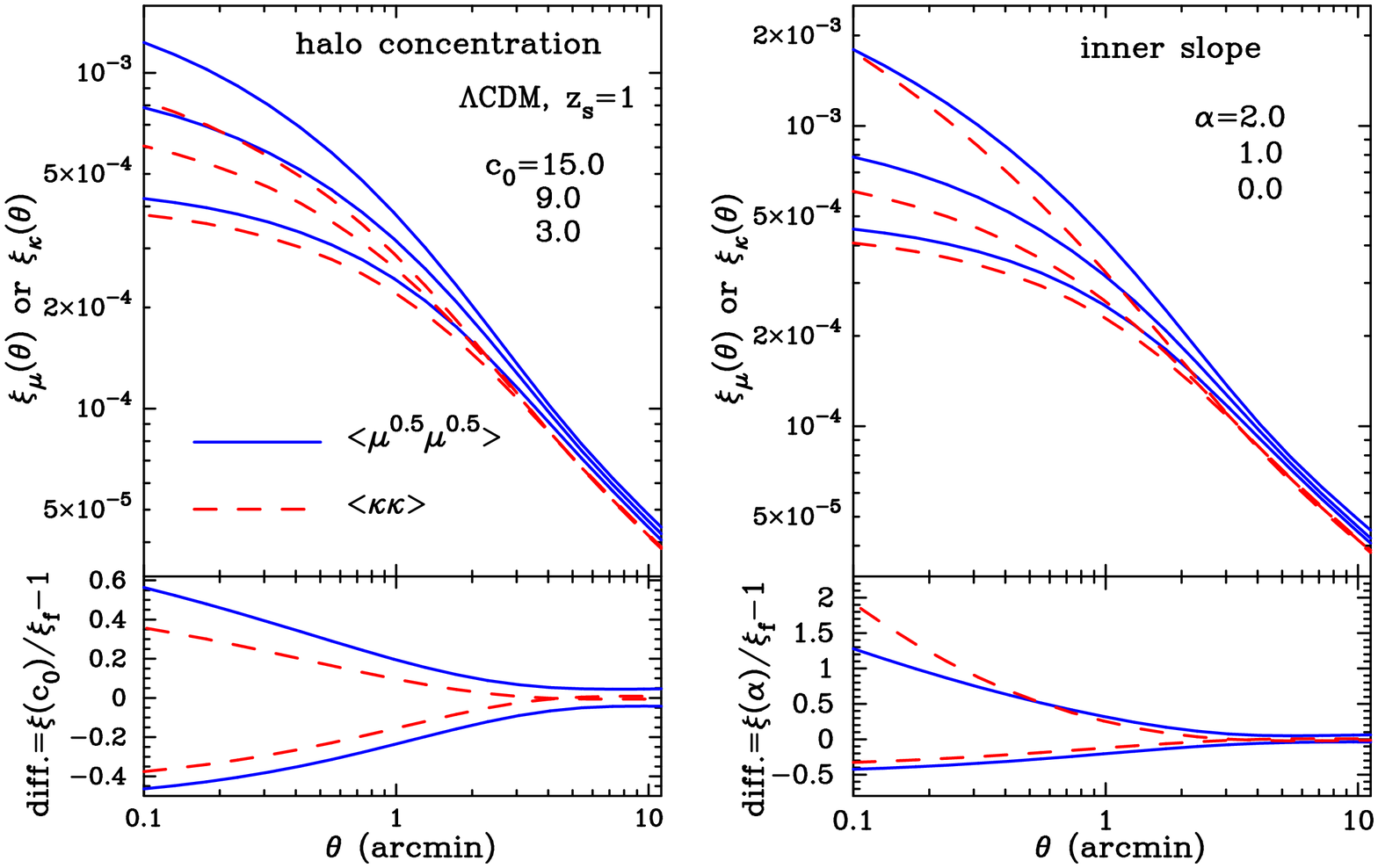}
  \end{center}
\caption{The dependences of the magnification 2PCF,
$\skaco{\delta\!\mu^{0.5}\cdot\delta\!\mu^{0.5}}$, on the halo profile
parameters.  The left panel shows the halo model predictions for the
halo concentration parameter $c_0=3, 9$ and $15$ (from bottom to top),
while the right panel shows the results with $\alpha=0$, $1$ and $2$ for
the inner slope parameter of the generalized NFW profile (see text for
more details). For comparison, the dashed curves are the corresponding
results for the convergence 2PCF. Note that the magnification 2PCF
considered becomes equivalent to the convergence 2PCF in the weak
lensing limit as $\delta\!\mu^{0.5}\approx \kappa$.  The lower panel
shows the difference relative to the result with our fiducial model of
$\alpha=1$ and $c_0=9$.  }\label{fig:alpha}
\end{figure*}
As discussed above, one of the useful cosmological information extracted
from the QSO-galaxy correlation measurement is information on the halo occupation
number of galaxies, which in turn provides a clue to understanding of
galaxy formation in connection to the dark matter halo properties (see
Jain et al. 2003 for the details). We here demonstrate another
possibility of the magnification statistics (especially measured via
galaxy size fluctuations) to address questions: what can we learn from
the measurements?  How is this method complementary to the established
cosmic shear that probes the correlations of the convergence
or shear fields ($\kappa$ or $\gamma$)?  To examine this, we focus on
the non-linear relation between the magnification and the cosmic shear
fields, as given by equation (\ref{eqn:mag}).
The non-linear effect is more pronounced on smaller scales, as have so
far been shown.  Future massive surveys promise to measure the
magnification statistics as well as the cosmic shear even on
sub-arcminute scales (Jain 2002; TJ03c; Jain et al. 2003).  Within a
picture of the halo model, the sub-arcminute correlation function is
quite sensitive to the halo profile properties (TJ03b,c) and 
the measurement can be potentially used to constrain the properties, if
the systematics is well under control.  Hence, we here investigate the
dependence of the magnification 2PCF on the halo profile parameters, the
halo concentration and the inner slope of the generalized NFW profile.
These parameters are still uncertain observationally and theoretically
and have information on the dark matter nature as well as
properties of highly non-linear gravitational clustering on $\simlt
1$Mpc.  Following TJ03c, we consider the parameterization
 given as $c(M,z)=r_{\rm vir}/r_s
=c_0(M/M_\ast)^{-0.13}$ and $\rho(r)\propto
r^{-\alpha}(1+r/r_s)^{-3+\alpha}$, respectively. Our fiducial model so
far used is given by $(c_0,\alpha)=(9.0,1.0)$.  For cases $\alpha=0$,
$1$ and $2$ we can derive analytic expressions for the convergence and
shear profiles from which we can also compute the 
magnification profile (the expressions of the convergence fields are
given in Appendix B in TJ03c).  Note that in what follows we employ the
virial boundary condition.
The relevant angular
scales are below the angular resolution of $N$-body simulations 
we have used.

The left panel of Figure \ref{fig:alpha} shows the halo model prediction
for the magnification 2PCF with varying the halo concentration, while
the right panel shows the results with varying $\alpha$.  Here we
consider $\skaco{\delta\!\mu^{0.5}\cdot \delta\!\mu^{0.5}}$, because
 it is less sensitive to high magnification
events ($\mu\gg 1$) and therefore observationally more robust (see
Figure \ref{fig:mumax}).  
In the weak lensing limit, the
correlation can be approximated by the convergence 2PCF (dashed curves),
which is measured by the cosmic shear measurement
because $\delta\mu^{0.5}\approx \kappa$.  Therefore, the
difference between the solid and dashed curves
reflects contribution from the non-linear magnifications $\delta
\mu\simgt 1$.  Halos with
masses $M\ge 10^{13}M_\odot$ provide dominant contribution of $\simgt
80\%$ to the total power over a range of non-linear scales $0.\!\!'
1$-$3'$ (e.g., see Figure 14 in TJ03c).  One can see that the
magnification 2PCF has stronger sensitivity to the halo concentration
and depends on the inner slope in a different way from the convergence
2PCF.
In TJ0c,
it was pointed out that the cosmic shear measurement introduces a
degeneracy in determining these halo profile parameters, even provided
the accurate measurement (see Figures 16 and 17 in TJ03c).  
The
results in Figure \ref{fig:alpha} thus indicate that a joint
measurement of the magnification statistics and the cosmic shear can be
used to improve the parameter determinations. 
Finally, one caution we make is 
that the magnification 2PCF for the profile with
$\alpha=2$ is more amplified by an increase of the maximum
magnification cutoff $\mu_{\rm max}$ than the other $\alpha$'s and thus
is sensitive to the selection effect.

\section{Discussions}
\label{disc}

In this paper, we have used the real-space halo approach to compute the
magnification correlation function without employing the weak lensing
approximation $\mu\approx 1+2\kappa+O(\kappa^2)$. 
It has been shown that the correction due to the non-linear
magnification
($\delta\mu\simgt
1$) leads to significant enhancement in the correlation amplitude
relative to the weak lensing approximation (see Figures
\ref{fig:compzs1}-\ref{fig:gbias}).
  The correction is more
important as one considers the correlation function for sources at
higher redshifts and on smaller angular scales, where the weak lensing
approximation ceases to be accurate. 
Thus, accounting for the non-linear
contribution in the theoretical model is needed to extract unbiased,
cosmological information from 
the precise measurement
expected from forthcoming and future surveys.
The encouraging result
shown is the halo model prediction remarkably well reproduces the
simulation result over the angular scales we consider.

We also developed the model to predict the QSO-galaxy cross-correlation
by incorporating the realistic model of the halo occupation number of
galaxies into the halo model (see \S \ref{qso}). The primary
cosmological information provided from the measurement is constraints on
the halo occupation number, as shown in Jain et al. (2003; also
see Guzik \& Seljak 2002). In particular, the QSO-galaxy correlation can
be used to directly
constrain the first moment of the halo occupation number, compared to the
two-point correlation of galaxies that probes the second moment. 
 Exploring the halo occupation is compelling in that it
provides useful information on
the galaxy formation and the merging history in connection with the dark
matter halo properties. We showed that the non-linear
magnification amplifies the cross-correlation amplitude by
$10-25\%$ on arcminute scales.  The method of this
paper therefore 
provides the accurate model prediction that accounts for both the
non-linear magnification correction and the realistic galaxy bias. 

We found that the magnification statistics can be used to extract
cosmological information complementary to that provided from the cosmic
shear measurement.  We have demonstrated that the joint measurement on
angular scales $\simlt 3'$ could be used to precisely constrain the halo
profile properties (see Figure \ref{fig:alpha}).  
This possibility would
open a new direction in using the magnification statistics as a
cosmological probe beyond determination of fundamental cosmological
parameters (Bartelmann 1995; Bartelmann \& Schneider 2001; M\'enard \&
Bartelmann 2002; M\'enard et al. 2003).  Exploring the halo profile
properties with gravitational lensing will be a direct test of the CDM
scenario in the highly non-linear regime $\simlt $Mpc, since alternative
scenarios have been proposed in order to reconcile the possible
conflicts between the CDM predictions and the observations on the small
scales (e.g., Spergel \& Steinhardt 2000).

In most results shown, we employed the maximum magnification cutoff
$\mu_{\rm max}=8$ for the halo model predictions as well as for the
simulation results, 
because the choice likely removes strong lensing
events ($\mu\simgt 10$) from the analysis. 
  Even if we employ the smaller value,
the qualitative conclusions derived  are not largely
changed, as can be seen from Figure \ref{fig:mumax}. 
Observationally, strong
lensing event is easily removed from the sample of the magnification 
statistics, if it accompanies multiple
images or largely deformed images. 
However, without the clear signature, 
it is relatively difficult to make a clear discrimination 
of the strong lensing, 
since the magnification is not a direct observable.
One advantage of the halo model developed in this paper is 
that it allows a fair comparison with the measurement 
by employing the selection criteria in the measurement for the model
prediction. Based on these considerations, 
we derived useful, general dependences of the
magnification correlation amplitude on large magnifications $(\mu\gg 1)$, from
the universal lensing properties in the vicinity of critical curves (see
\S \ref{asymp}).  The intriguing consequence is that, for a correlation
function parameterized as $\skaco{\mu^p f}$, the amplitude converges to
be finite for $p<1$ and otherwise diverges $p\ge 1$ as the maximum
magnification cutoff $\mu_{\rm
max}\rightarrow \infty$, {\rm independent} of details of the lensing
mass distribution. This was quantitatively verified by the halo model
prediction as well as by the simulations (see Figure \ref{fig:mumax}).
This result therefore implies
that the magnification statistics with
$p\le 1$ are practically advantageous in that it is
insensitive to the selection effect of the
magnification cutoff $\mu_{\rm max}$.  This is the case for
the two-point correlation of size (not area) fluctuations of distant
galaxy images and for the QSO-galaxy cross correlation, if the unlensed
number counts of QSOs with a limiting magnitude have a slope of $s=d\ln
N(m)/dm<4/5$. 

One might imagine that the non-linear magnification contribution can be
suppressed by clipping regions of cluster of galaxies from survey data
in order to avoid the uncertainty in the model prediction and to apply
the weak lensing approximation (e.g., see
 Barber \& Taylor 2002). 
However, this likely adds
an artificial selection effect in the analysis and causes a biased 
cosmological interpretation.
In addition, the lensing
projection makes it relatively difficult to correctly identify the
cluster region from the reconstructed convergence map, unless
accurate photo-$z$ information or follow-up observations are
available (e.g., White et al. 2002).  The approach of this paper allows
us to treat data more objectively. 

There are some effects we have so far ignored in the halo model
calculation. Most important one is the asphericity of halo profile in a
statistical sense.  The aspherical effect could lead to substantial
enhancement of the magnification correlation amplitude, since it is
known that an area enclosed by the critical curve is largely increased
by ellipticity of the lensing mass distribution, thus increasing the
cross section of high magnifications to the correlation evaluation.  For
example, the number of strong lensing arcs due to clusters of galaxies
is amplified by an order of magnitudes if one considers an elliptical
lens model instead of an axially symmetric profile (e.g., Meneghetti et
al. 2003; Oguri et al. 2003). For the same reason, substructures within
a halo could have strong impact on the magnification correlation, as
they naturally emerges in the CDM simulations.  Hence, it is of great
interest to investigate the effect on the magnification statistics 
with higher resolution simulations.

\bigskip

We are grateful to Bhuvnesh Jain for collaborative work, valuable
discussions and a careful reading of the manuscript.  
T.~H. would like to
thank for his warm hospitality at University of Pennsylvania, where a
part of this work was done.  
We also thank M. Bartelmann and M. Oguri for valuable discussions.
 T.H. also acknowledges supports from Japan Society for
Promotion of Science (JSPS) Research Fellowships. 
Numerical computations presented in this paper were partly
carried out at ADAC (the Astronomical Data Analysis Center) of the
National Astronomical Observatory, Japan and at the computing centre of the
Max-Planck Society in Garching.
The $N$-body simulations used in this paper were carried out by the Virgo
Supercomputing Consortium.

\label{lastpage}
\end{document}